\documentclass[twocolumn,showpacs,preprintnumbers,amssymb,pra]{revtex4}
\usepackage{latexsym,epsfig}
\usepackage{graphicx, color}
\usepackage{dcolumn}
\usepackage{amsthm,amsmath}

\begin{document}
\title{Relativistic coupled-cluster analysis of parity nonconserving amplitudes and related properties of the $6s ~ ^2S_{1/2} - 5d ~ ^2D_{3/2;5/2}$ 
transitions in $^{133}$Cs} 

\author{B. K. Sahoo \footnote{Email: bijaya@prl.res.in}}
\affiliation{Atomic, Molecular and Optical Physics Division, Physical Research Laboratory, Ahmedabad-380009, India}

\author{B. P. Das \footnote{Email: bpdas.iia@gmail.com}}
\affiliation{Department of Physics International Education and Research Center of Science, Tokyo Institute of Technology,
2-1-2-1-H86 Ookayama Meguro-ku, Tokyo 152-8550, Japan}

\date{Received date; Accepted date}

\begin{abstract}
We present the results of our calculations of the parity nonconserving electric dipole amplitudes ($E1_{PNC}$) for the $6s ~ ^2S_{1/2} - 5d ~ ^2D_{3/2;5/2}$ 
transitions of $^{133}$Cs employing a relativistic coupled-cluster (RCC) method. $E1_{PNC}$ values for the nuclear spin independent (NSI) and nuclear spin
dependent (NSD) parity non-conservation (PNC) effects for  different hyperfine levels of the $6s ~ ^2S_{1/2} - 5d ~ ^2D_{3/2}$ transition 
and only due to the NSD PNC interaction Hamiltonian for all possible hyperfine levels of the $6s ~ ^2S_{1/2} - 5d ~ ^2D_{5/2}$ transition 
are given. We have also performed calculations of several properties that are relevant for assessing the reliability of the 
calculations of the amplitudes of the above mentioned transitions. We highlight the importance of correlation effects by giving the
results of both the Dirac-Fock (DF) and RCC contributions. Approximate contributions to these properties from the Breit and quantum 
electrodynamics (QED) interactions have also been evaluated. Plausible experimental schemes to measure these transition amplitudes by 
light shift techniques and interference involving induced Stark shifts are outlined.
\end{abstract}



\maketitle

\section{Introduction}

It is possible to probe new physics by investigating parity nonconservation (PNC) effects in atomic systems 
\cite{bouchiat,commins,ginges}. The two sources of PNC interactions in atomic systems are  the neutral current weak interaction
due to the exchange of the Z$_0$ boson between the nucleons and electrons \cite{bouchiat1,erler,marciano} and the electromagnetic
 interaction between the electrons and a possibly existing nuclear anapole moment (NAM) \cite{zeldovich,flambaum,flam1}. 
Weak interactions due to exchange of Z$_0$ boson can be again classified into nuclear spin independent (NSI) and nuclear spin 
dependent (NSD) interactions depending upon whether the axial-vector and vector currents come from the electron and nuclear 
sectors respectively, or the other-way-around \cite{ginges}. These interactions, especially the NSI component owing to the coherent 
contributions from all the nucleons, are sensitive to probe possible new physics beyond the Standard Model (SM) of particle interactions
by determining the electron-up and electron-down quarks coupling coefficients \cite{bouchiat,bouchiat1}. The expected new physics from 
these couplings can also be studied with the upgraded Large Hadron Collider (LHC) when it attempts to search for physics 
at the TeV energy scale. The concept of NAM is very fundamental, but its existence is still an open question \cite{ginges,zeldovich,flambaum}. 
Though there has been a claim that NAM has been observed in PNC measurements in atomic Cs \cite{wood}, the extracted nuclear coupling 
parameter inferred from this experiment does not agree with the  values obtained from nuclear data \cite{ginges,wilburn,haxton}. 
Moreover, the sign of the NAM coupling constant from the Cs measurement is not in agreement with  that from the Tl PNC measurement \cite{ginges}. To 
confirm these results, it is imperative to carry out further investigations on NAM, which can be achieved by studying PNC in more atomic 
systems or considering other suitable transitions in the Cs atom. In fact, a number of atoms such as Rb \cite{sheng}, Fr \cite{sheng,gomez1,bijaya1} 
and Yb \cite{tsigutkin} and singly charged ions like Ba$^+$ \cite{geetha,bijaya2,dzuba}, Ra$^+$ \cite{bijaya2,dzuba,wansbeek}, 
Yb$^+$ \cite{dzuba,roberts} etc. have been considered for PNC studies, partly to probe the existence of the NAM.

The PNC interaction due to the NAM is also NSD in character and contributes more in heavy atomic systems than the NSD 
PNC interaction mediated by the Z$_0$ boson \cite{ginges}. In many of the above mentioned systems, the dominant  contribution to the PNC 
effects in the transitions come from the  NSI PNC interaction. It is desirable for the NSI contributions to cancel to facilitate the 
observation of the NSD interaction which is dominated by the NAM in heavy atoms. This amounts to performing measurements more than once in different configurations. 
In contrast, if the measurement would be carried out for a transition in a heavy atom in which the PNC contribution is entirely due to the NSD interactions then that would be
an unambiguous signature of the nuclear anapole moment. Keeping this in mind the transition between the ground and the metastable 
$(n-1)d ~ ^2D_{5/2}$ states, where $n$ is the principal quantum number associated with the ground state of the alkaline earth ions
such as Ba$^+$ and Ra$^+$ were proposed to measure NSD PNC in these transitions \cite{geetha,bijaya1}. Although the NSD PNC interaction
due to the neutral weak current can contribute and the interference of the NSI PNC and the hyperfine interactions can also do so, 
albeit to an even smaller extent \cite{ginges}, 
the largest contribution would come from the NAM. The other advantages of these transitions are the long lifetimes of the metastable 
$(n-1)d ~ ^2D_{5/2}$ states in these ions and a novel technique of measuring PNC induced light shift resulting from the 
interference of the electric dipole ($E1_{PNC}$) amplitude of the forbidden $6s ~ ^2S_{1/2} - 5d ~ ^2D_{3/2}$ transition 
and the electric quadrupole (E2) transition amplitude in $^{137}$Ba$^+$, that was proposed by Fortson \cite{fortson}, can be employed in this case for carrying out such measurements. Recently, 
we had argued that a similar scheme with appropriate modifications can also be adopted to measure PNC amplitudes in the $7s ~ ^2S_{1/2} - 
6d ~ ^2D_{3/2;5/2}$ transitions in atomic Fr \cite{bijaya1}. Since the lifetimes of the $6d ~ ^2D_{3/2;5/2}$ states of Fr 
have been predicted to be smaller than the metastable $(n-1)d ~ ^2D_{5/2}$ states of alkaline-earth ions, precise
shot-noise-limits can be achieved by considering a large number of $(\sim 10^4)$ of Fr atoms in an optical lattice for these measurements 
\cite{bijaya1,aoki}. Cs atom, which also belongs to the alkali group, has an energy level structure similar to the Fr atom and the lifetimes of
its $5d ~ ^2D_{3/2;5/2}$ states are of the same order of magnitude as those of the $6d ~ ^2D_{3/2;5/2}$ states of Fr \cite{bijaya3,bijaya4}. Moreover, many high precision measurements related to atomic clocks, PNC 
and EDM, cold atom experiments etc. have already been performed on the $^{133}$Cs isotope \cite{wood,murthy,mckeever}. In this context, 
consideration of the $6s ~ ^2S_{1/2} - 5d ~ ^2D_{3/2;5/2}$ transitions in the Cs atom for the measurements of the PNC 
effects related to the NAM would be of great interest. Particularly, it is important to undertake studies on the 
$6s ~ ^2S_{1/2} - 5d ~ ^2D_{5/2}$ transition for this purpose. In fact, earlier studies had demonstrated \cite{ginges,roberts} that PNC amplitudes in the 
$6s ~ ^2S_{1/2} - 5d ~ ^2D_{3/2}$ transition is larger than the $6s ~ ^2S_{1/2} - 7s ~ ^2S_{1/2}$ transition where the most precise PNC measurement is 
currently available. The reason for this that the contributions of the $6s ~ ^2S_{1/2}$ and $7s ~ ^2S_{1/2}$ states to the PNC amplitude 
of the $6s ~ ^2S_{1/2} - 7s ~ ^2S_{1/2}$ transition are substantial, but their signs  are opposite, resulting in a large cancellation to 
give the final result \cite{ginges,dzuba,roberts,bijaya1,shabaev,derevianko,bijaya5,bijaya6}. In contrast, only the $6s ~ ^2S_{1/2}$
state is the dominant contributor to the PNC amplitude of the $6s ~ ^2S_{1/2} - 5d ~ ^2D_{3/2}$ transition. In this work, we have carried out a 
relativistic coupled-cluster (RCC) theory analysis of the PNC amplitudes in the $6s ~ ^2S_{1/2} - 5d ~ ^2D_{3/2;5/2}$ transitions of $^{133}$Cs
and demonstrate their enhanced values compared to the previously reported values due to certain all order many-body effects. Also, it was 
observed recently from the comparative study of theoretical and experimental values of the lifetimes of the $5d ~ ^2D_{3/2;5/2}$ states 
in Cs that correlation effects that contribute through the non-linear terms of RCC theory are crucial for determining the 
$5d ~ ^2D_{3/2;5/2}$ states \cite{bijaya3}.
Thus, it is necessary to analyze the PNC amplitudes of the $6s ~ ^2S_{1/2} - 5d ~ ^2D_{3/2;5/2}$ transitions in $^{133}$Cs using the full fledged
RCC theory. In addition, we discuss the feasibility of measuring PNC observables either using the techniques of light-shift 
detection or using a  technique similar to the Stark induced electric dipole ($E1_{Stark}$) amplitude measurement as had been employed in 
the case of the $6s ~ ^2S_{1/2} - 7s ~ ^2S_{1/2}$ transition of $^{133}$Cs \cite{wood}.

\section{Theory}
 
The atomic Hamiltonian due to the NSI PNC interaction is given by \cite{ginges}
\begin{eqnarray}
\label{eq1}
H_{PNC}^{NSI}&=&-\frac{G_F}{2\sqrt{2}}Q_{W} \gamma_{5} \rho_{n}(r),
\end{eqnarray}
where $G_F$ is the Fermi constant, $Q_W$ is the weak charge, $\rho_{nuc}$ is the nuclear density, and  $\gamma_{5}$ is 
the Dirac matrix. Similarly, the Hamiltonian due to the NSD PNC interaction is given by \cite{ginges}
\begin{eqnarray}
H_{PNC}^{NSD} &=& \frac {G_F}{\sqrt{2}} \frac{K_W}{I} \mbox{\boldmath$\alpha$} \cdot {\bf I} \ \rho_{n}(r) \nonumber \\
              &=& \sum_q (-1)^q I_q^{(1)} K_{-q}^{(1)},
\label{eq2}
\end{eqnarray}
where  $\mbox{\boldmath$\alpha$}$ is the Dirac matrix, $I=|{\bf I}|$ is the nuclear spin with its $q^{th}$ component $I_q$, the 
dimensionless quantity $K_W$ is related to NAM and electron-quark coefficients due to NSD PNC interactions and $K_{q}^{(1)}$ is 
the $q^{th}$ component of a tensor ${\bf K}$ containing electronic component of $H_{PNC}^{NSD}$. Therefore, the 
combined net PNC interaction Hamiltonian in an atomic system is given by
\begin{eqnarray}
 H_{PNC} &=& H_{PNC}^{NSI} + H_{PNC}^{NSD} \nonumber \\
         &=& \frac{G_F}{\sqrt{2}} \left ( - \frac{Q_{W}}{2} \gamma_{5} + \frac{K_W}{I}  \mbox{\boldmath$\alpha$} \cdot {\bf I} \right ) \ \rho_{nuc}(r) .
\end{eqnarray}

Consideration of the $H_{PNC}$ interaction Hamiltonian will produce orbitals of mixed parity. As a result, parity forbidden transitions in the atomic systems 
can have small but finite 
probabilities with amplitudes $E1_{PNC}$. The electric dipole $E1_{PNC}^{M_f M_i} $ amplitude due to these 
PNC interactions between the hyperfine states $|F_f,M_f \rangle$ and $|F_i,M_i \rangle$ can be expressed as
\begin{eqnarray}
E1_{PNC}^{M_f M_i} &=& (-1)^{F_f - M_f} \left ( \begin{matrix} F_f & 1 & F_i \cr -M_f & q & M_i \cr \end{matrix} \right ) \left (\mathcal{X+Y} \right ),
\label{eq4}
\end{eqnarray}
where $q=-1$, 0 or 1 depending on the choice of the $M$-values. In the above expression, $\mathcal{X}$ and $\mathcal{Y}$ 
correspond to the reduced matrix elements for the contributions from the NSI and NSD components, respectively, and are
expressed as
\begin{eqnarray}
 \mathcal{X} &=& (-1)^{I+F_i + J_f +1 } \sqrt{(2F_f+1)(2F_i+1)}   \left \{ \begin{matrix} J_i & J_f & 1 \cr F_f & F_i & I \cr \end{matrix} \right \}  
 \nonumber \\ && \times \Big ( \sum_{k\ne i} \frac{ \langle J_f || {\bf D} || J_k \rangle \langle J_k || {\bf H}_{PNC}^{NSI} || J_i \rangle}{ E_i - E_k} \nonumber \\  
&& + \sum_{k\ne f} \frac{ \langle J_f || {\bf H}_{PNC}^{NSI} || J_k \rangle \langle J_k || {\bf D} || J_i \rangle}{ E_f - E_k}  \Big )
\label{eq5}
\end{eqnarray}
and
\begin{eqnarray}
\mathcal{Y} &=& \eta \ \Big ( \sum_{k\ne i} (-1)^{j_i - j_f +1} \frac{ \langle J_f || {\bf D} || J_k \rangle \langle J_k || {\bf K} || J_i \rangle}{ E_i - E_k} \nonumber \\ 
&& \times  \left \{ \begin{matrix} F_f & F_i & 1 \cr J_k & J_f & I \cr \end{matrix} \right \} \left \{ \begin{matrix} I & I & 1 \cr J_k & J_i & F_i \end{matrix} \right \} \nonumber \\ 
&& + \sum_{k\ne f} (-1)^{F_i - F_f +1} \frac{ \langle J_f || {\bf K} || J_k \rangle \langle J_k || {\bf D} || J_i \rangle}{ E_f - E_k} \nonumber \\
&&  \times \left \{ \begin{matrix} F_f & F_i & 1 \cr J_i & J_k & I \cr \end{matrix} \right \} \left \{ \begin{matrix} I & I & 1 \cr J_k & J_f & F_f \cr \end{matrix} \right \} \Big ),
\label{eq6}
\end{eqnarray}
where $\eta=\sqrt{(I+1)(2I+1) (2F_i+1) (2F_f+1)/I}$ and $J$ and $E$s are the total angular momentum and energies of the 
respective states. The index $k$ runs over allowed intermediate states.

\section{Method of calculations}

We consider the Dirac-Coulomb (DC) Hamiltonian to account for the relativistic and electron correlation effects in the RCC method for
calculating atomic wave functions. In atomic units (a.u.), it is given by
{\small 
\begin{eqnarray}
H_{DC} &=& \sum_i \left [ c\mbox{\boldmath$\alpha$}_i\cdot \textbf{p}_i+(\beta_i -1)c^2 + V_n(r_i) + \sum_{j>i} V_C(r_{ij}) \right ] 
\end{eqnarray}}
with $\mbox{\boldmath$\alpha$}$ and $\beta$ are the usual Dirac matrices, $V_n(r)$ represents the nuclear potential and 
$V_C(r_{ij})=\frac{1}{r_{ij}}$ is the Coulomb interaction potential between the electrons. For evaluating $\rho_n(r)$ and 
$V_n(r)$, we consider the Fermi-charge distribution defined by
\begin{equation}
\rho_{n}(r_i)=\frac{\rho_{0}}{1+e^{(r_i-b)/a}} 
\end{equation}
for the normalization factor $\rho_0$, the half-charge radius $b$ and $a= 2.3/4(ln3)$ is related to the skin thickness. We 
have determined $b$ using the relation
\begin{eqnarray}
b&=& \sqrt{\frac {5}{3} r_{rms}^2 - \frac {7}{3} a^2 \pi^2}
\end{eqnarray}
with the root mean square (rms) charge radius of the nucleus evaluated by using the formula
\begin{eqnarray}
 r_{rms} =0.836 A^{1/3} + 0.570
\end{eqnarray}
in $fm$ for the atomic mass $A$.

We also account contributions from the transverse photons by adding the Breit interaction to the DC Hamiltonian, given by
\begin{eqnarray}
V_B(r_{ij})&=& - \frac{\{\mbox{\boldmath$\alpha$}_i\cdot \mbox{\boldmath$\alpha$}_j+
(\mbox{\boldmath$\alpha$}_i\cdot\bf{\hat{r}_{ij}})(\mbox{\boldmath$\alpha$}_j \cdot \bf{\hat{r}_{ij}}) \}}{r_{ij}}, \ \ \ \ \ \
\end{eqnarray}
along with the Coulomb potential $V_C(r_{ij})$, where $\bf{\hat{r}_{ij}}$ is the unit vector along the inter-electronic 
distance. Similarly, corrections from the lower order quantum electrodynamic (QED) effects 
are estimated by considering effective potentials along with the nuclear potential $V_n(r_i)$ in the atomic Hamiltonian 
$H_{DC}$ that are given by
\begin{eqnarray}
V_{QED}(r_i) &=& V_U(r_i) +  V_{WK}(r_i) + V_{SE}^{ef}(r_i) + V_{SE}^{mg}(r_i) , \ \ \ \ \
\end{eqnarray}
where $V_{U}(r)$ and $V_{WK}(r)$ are known as Uehling and Wichmann-Kroll potentials, respectively, that account lower 
order vacuum polarization (VP) effects and $V_{SE}^{ef}(r)$ and $V_{SE}^{mg}(r)$ are the electric and magnetic form-factor 
contributions from the lower-order self-energy (SE) effects, respectively. In our previous work on the Cs atom \cite{bijaya4}, we have
given these expressions with a Fermi charge distribution. 

For computational simplicity, we obtain first the Dirac-Fock (DF) wave function ($\vert \Phi_0 \rangle$) using the aforementioned interaction 
Hamiltonians, without PNC interactions, for the $[5p^6]$ closed-shell configuration of the Cs atom. The single particle orbitals of 
this DF wave function are defined as 
\begin{eqnarray}
 |\varphi_{n,\kappa,m} (r,\theta,\phi)=\frac{1}{r} \left ( \begin{matrix} P_{n,\kappa}(r) & \chi_{\kappa,m}(\theta,\phi) \cr i Q_{n,\kappa}(r) & \chi_{-\kappa,m}(\theta,\phi) \cr \end{matrix} \right ),
\end{eqnarray}
where $P_{n,\kappa}(r)$ and $Q_{n,\kappa}(r)$ are the large and small component radial components with $\chi_{\kappa,m}(\theta,\phi)$ and
$\chi_{-\kappa,m}(\theta,\phi)$ are the respective four component spinors for the principal quantum number $n$ and relativistic angular 
momentum quantum number $\kappa$. The radial components are constructed using Gaussian type orbitals (GTOs), that is 
\begin{eqnarray}
 P_{n,\kappa}(r) = \sum_{k=1}^{N_k} C_{\kappa,k}^L g_{\kappa,k}^L(r)
\end{eqnarray}
and
\begin{eqnarray}
 Q_{n,\kappa}(r) = \sum_{k=1}^{N_k} C_{\kappa,k}^S g_{\kappa,k}^S(r),
\end{eqnarray}
where $k$ sums over the total number of GTOs ($N_k$) in each symmetry, $C_{\kappa,k}^{L/S}$ are the unknown coefficients that need to 
be determined and $g_{\kappa,k}^{L/S}$ are the GTOs for the large ($L$) and small ($S$)
components respectively. The GTOs for the large radial component are defined as
\begin{eqnarray}
g_{\kappa,k}^L(r)  &=&  {\cal N}_{\kappa,k}^L  r^{l+1} e^{-(\eta_k r^2)},
\label{anbas}
\end{eqnarray}
where ${\cal N}_{\kappa,k}^L$ represents for normalization constant, $\eta_k$ is an arbitrary coefficients suitably chosen for accurate 
calculations of wave functions and $l$ is the orbital quantum number of the orbital. The exponents $\eta_k$ form an even-tempered 
series
\begin{eqnarray}
 \eta_k = \eta_0 \zeta^{k-1}
\end{eqnarray}
in terms of the parameters $\eta_0$ and $\zeta$. However, GTOs for the small radial component are defined 
by implementing kinetic balance condition as
\begin{eqnarray}
g_{\kappa,k}^S(r)  &=&  {\cal N}_{\kappa,k}^S \left ( \frac{d}{dr} + \frac{\kappa}{r} \right ) g_{\kappa,k}^L(r),
\end{eqnarray}
with the corresponding normalization constant ${\cal N}_{\kappa,k}^S$. We give the list of $\eta_0$ and $\zeta$ parameters along with
$N_k$ for each symmetry in Table \ref{tab1} that are used in the present calculations.

\begin{table}[t]
\caption{List of number of GTOs and $\eta_0$ and $\zeta$ parameters used to define the basis functions for different symmetries 
to construct single particle orbitals in the present calculations.}
 \begin{ruledtabular}
  \begin{tabular}{lccccc}
   &  $s$ & $p$ & $d$& $f$ & $g$  \\
  \hline
 & & & \\
 $N_k$  & 34 & 33 & 32 & 31 & 30 \\
 $\eta_0$ & $0.0005$ &  $0.0015$ & $0.0035$ & $0.0051$ & $0.0071$ \\
 $\zeta$ & 2.01 & 1.96 & 2.01 & 1.96 & 2.01 \\
  \end{tabular}
 \end{ruledtabular}
 \label{tab1}
\end{table}

The important intermediate states including the initial and final states of the considered transitions for PNC studies in $^{133}$Cs
have the common inert core, i.e. $[5p^6]$, of Xe atom (configuration of Cs$^+$) and a valence electron. Conveniently we obtain the single 
particle orbitals of $\vert \Phi_0 \rangle$ for the closed-core using the DF method and generate configurations for the respective 
open-shells by appending a corresponding virtual orbital as the valence orbital. In this formalism, both $\mathcal{X}$ and $\mathcal{Y}$ 
can be evaluated employing a sum-over-states approach by dividing total contributions to these quantities into two parts as
\begin{eqnarray}
\mathcal{X} &=& \mathcal{X}^{v} + \mathcal{X}^{cv}
\end{eqnarray}
and
\begin{eqnarray}
\mathcal{Y} &=& \mathcal{Y}^{v} + \mathcal{Y}^{cv},
\end{eqnarray}
where the superscripts $v$ and $cv$ correspond to contributions from the electron ``correlations'' involving the valence orbital and 
``relaxation'' effects to the core-orbitals due to the valence orbital that were neglected in the construction of core-orbitals in the
DF method, respectively. The ``relaxation'' contributions are extremely small and are estimated at the DF level, thus 
without accounting for electron correlation contributions. The valence correlation effects are determined by dividing further into two parts: the
contributions from the low-lying bound states and the contributions from high-lying states including the continuum. The 
first part is referred to as ``Main'' contribution while the later part is referred to as the ``Tail'' contributions to $\mathcal{X}^v$ and 
$\mathcal{Y}^v$. We intend to consider contributions from as many as low-lying states to ``Main'' by evaluating the required matrix 
elements of the dipole and PNC interaction operators using a RCC theory as described below. The non-significant ``Tail'' contributions 
are again estimated using the DF method.

The ``Main'' contributions to the valence correlation effects are determined by evaluating contributions from as many as 
intermediate excited states possible. We calculate these intermediate states along with the initial and final unperturbed states
using the exponential {\it ansatz} in the RCC theory framework, in which a state with a closed-core and a valence orbital $v$ is
expressed as
\begin{eqnarray}
 \vert \Psi_v \rangle  = e^T \{ 1+ S_v \} \vert \Phi_v \rangle ,
 \label{eqcc}
\end{eqnarray}
where $\vert \Phi_v \rangle=a_v^{\dagger} \vert \Phi_0 \rangle$. Here, $T$ and $S_v$ are the RCC excitation operators that excite electrons from $\vert \Phi_0 \rangle$
and $\vert \Phi_v \rangle$, respectively, to the virtual space. It can be noted that the above expression is linear in $S_v$ operator 
owing to presence of only one valence orbital $v$. This expression is, however, exact and it accounts for the non-linear effects through  
the products of $T$ and $S_v$ operators. In this work, we have considered only the single and double excitations in the RCC theory in 
the CCSD method approximation by expressing $T=T_1+T_2$ and $S_v=S_{1v}+S_{2v}$. The amplitudes of these RCC operators are evaluated 
by solving the following coupled-equations for the singles and doubles excitations in the Jacobi iterative procedure
\begin{eqnarray}
 \langle \Phi_0^* \vert \overline{H}  \vert \Phi_0 \rangle &=& 0
\label{eqt}
 \end{eqnarray}
and
\begin{eqnarray}
 \langle \Phi_v^* \vert \big ( \overline{H} - \Delta E_v \big ) S_v \vert \Phi_v \rangle &=&  - \langle \Phi_v^* \vert \overline{H}_N \vert \Phi_v \rangle ,
\label{eqsv}
 \end{eqnarray}
where $\vert \Phi_0^* \rangle$ and $\vert \Phi_v^* \rangle$ are the excited state configurations, here up to doubles, with
respect to the DF states $\vert \Phi_0 \rangle$ and $\vert \Phi_v \rangle$ respectively. Here $\overline{H}= \big ( H e^T
\big )_l$ with subscript $l$ represents for the linked terms only and $\Delta E_v$ is the electron attachment energy (EA) 
of the electron of the valence orbital $v$. We evaluate $\Delta E_v$ by
\begin{eqnarray}
 \Delta E_v  = \langle \Phi_v \vert \overline{H} \left \{ 1+S_v \right \} \vert \Phi_v \rangle - \langle \Phi_0 | \overline{H} | \Phi_0 \rangle .
 \label{eqeng}
\end{eqnarray}
Both Eqs. (\ref{eqsv}) and (\ref{eqeng}) are solved simultaneously, as a result Eq. (\ref{eqsv}) effectively becomes non-linear in
the $S_v$ operator. In fact, the excitation energy (EE) between two given states is evaluated by taking difference between the 
respective EAs obtained from the above procedure.

\begin{table*}[t]
\caption{Estimated $E1_{PNC}$ amplitudes among possible hyperfine levels of the $6s ~ ^2S_{1/2} (F_i ) - 5d ~ ^2D_{3/2} (F_f )$ and 
$6s ~ ^2S_{1/2} (F_i ) - 5d ~ ^2D_{5/2} (F_f )$ transitions of $^{133}$Cs due to NSI PNC (given as $E1_{PNC}^{NSI}$) and NSD PNC 
(given as $E1_{PNC}^{NSD}$) interactions in unit of $iea_0 \times 10^{-11}$. We have used $Q_W=-73.2$ and $K_W=0.419$ to estimate
these quantities. Magnetic quantum number is chosen as the lower value among $F_i$ and $F_f$. We also compare our results with another 
calculation reported in Ref. \cite{roberts}.}
\begin{tabular}{ll |ccc|ccc| ccc}
\hline \hline
 \multicolumn{10}{c}{} \\
   &   &  \multicolumn{6}{|c|} {$5d ~ ^2D_{3/2} (F_f) - 6s ~ ^2S_{1/2} (F_i)$} & \multicolumn{3}{c}{$5d ~ ^2D_{5/2} (F_f) - 6s ~ ^2S_{1/2} (F_i)$} \\
 \cline{3-8} \cline{9-11} \\
  $F_f$      &  $F_i$      &  \multicolumn{3}{|c|}{$E1_{PNC}^{NSI}$}  & \multicolumn{3}{c|}{$E1_{PNC}^{NSD}$  ($\times 10^{-3}$)} & \multicolumn{3}{c}{$E1_{PNC}^{NSD}$  ($\times 10^{-4}$)} \\
\cline{3-5} \cline{6-8} \cline{9-11} \\
  &             &     DF  &   CCSD  & Ref. \cite{roberts} &  DF & CCSD & Ref. \cite{roberts}$^{\dagger}$ & DF & CCSD & Ref. \cite{roberts}$^{\dagger}$\\
\hline
 & & & & & & & \\
  2    &  3   & $-1.737$ & $-2.064$ & $-2.05$ & $-18.01$ & $-32.11$ & $-38.14$ & $-0.002$ & 82.68 & 31.13 \\
  3    &  3   & $-2.669$ & $-3.181$ & $-3.14$ & $-27.69$ & $-44.29$ &$-56.70$ &  $-0.004$ & 183.44 & 69.14 \\
  4    &  3   & 1.149 & $1.368$ & 1.35 & 11.92 & $15.04$ & 23.30 & 0.002 & $-90.69$ & $-34.15$ \\ 
  3    &  4   & $-0.785$ & $-0.936$ & $-0.923$ & 6.33 & 5.82 & 11.90 & $-0.001$ & 53.92 & 20.28 \\
  4    &  4   & $-2.431$ & $-2.898$ & $-2.86$ &  19.62 & $26.47$ & 38.71  & $-0.004$ & 191.96 & 72.07 \\
  5    &  4   & 1.592 & $1.897$ & 1.87 & $-12.84$ &  $-22.94$ & $-27.03$  & 0.002 & $-115.76$ & $-43.57$ \\
 \hline \hline                            
\end{tabular}
\label{tab2}\\
$^{\dagger}$ We have multiplied by $K_W=0.419$ for the the comparison.
 \end{table*}
After obtaining amplitudes of the RCC operators using the above described equations,, the transition matrix element
of an operator $O$ between the states $\vert \Psi_i \rangle$ and $\vert \Psi_f \rangle$ is evaluated using the expression
\begin{eqnarray}
\langle \Psi_f| O| \Psi_i\rangle &=& \frac{\langle\Phi_f|\tilde{O}_{fi}|\Phi_i\rangle}{\sqrt{\langle\Phi_f|\{1+\tilde{N}_f\}|\Phi_f\rangle
\langle\Phi_i|\{1+\tilde{N}_i\}|\Phi_i \rangle}} , \nonumber \\
\label{eqno}
\end{eqnarray}
where $\tilde{O}_{fi}=\{1+S_f^{\dagger} \} e^{T^{\dagger}} O e^T \{1+S_{i}\}$ and $\tilde{N}_{k=f,i}=\{1+S_k^{\dagger} \} e^{T^{\dagger}}e^T \{1+S_{k}\}$.
For the expectation value of an operator $O$, the same expression is employed setting $\vert \Psi_i \rangle = \vert \Psi_f \rangle$.
As can be seen, it involves two non-terminating series in the numerator and denominator in the above expression, which are
$e^{T^{\dagger}} O e^T$ and $e^{T^{\dagger}} e^T$ respectively. As described in our previous works \cite{bijaya1,bijaya3,bijaya4},
we adopt iterative procedures to account for contributions from these non-terminating series. In this procedure, we divide the above
expressions into effective one-body, two-body and three-body terms using Wick's generalized theorem \cite{bartlett}. The effective 
one-body terms are dominant and are computed self-consistently. After computing we store them as intermediate parts before contracting 
with the corresponding $S_i$ and $S_f^{\dagger}$ operators. To consider terms from the above non-terminating series systematically, we 
first consider the linear terms and then contract them with another $T$ or $T^{\dagger}$ operator one by one gradually in the iterative 
scheme till we achieve a difference of $10^{-8}$ between two successive operations. We also use these terms to construct the effective 
two-body and three-body terms, which are computed directly after contracting with the $S_i$ and $S_f^{\dagger}$ operators.

\section{Possible experimental schemes}

The most precise PNC measurement in the $6s ~ ^2S_{1/2} - 7s ~ ^2S_{1/2}$ transition of $^{133}$Cs was carried out by interfering the Stark induced electric
dipole amplitude ($E1_{Stark}$) with the $E1_{PNC}$ amplitudes \cite{wood}. The NSI and NSD contributions from these measurements were separated out by 
taking average or subtracting measured values of these inferences between the hyperfine states $6s ~ ^2S_{1/2}(F=3) - 7s ~ ^2S_{1/2}(F=4)$ 
and $6s ~ ^2S_{1/2} (F=4) - 7s ~ ^2S_{1/2} (F=3)$ transitions, respectively. We also propose to adopt a similar technique by 
measuring inferences between the $6s ~ ^2S_{1/2} (F=3) - 5d ~ ^2D_{3/2} (F=4)$ and $6s ~ ^2S_{1/2} (F=4)- 5d ~ ^2D_{3/2} (F=3)$ 
transitions to extract out both the NSI and NSD contributions in the $6s ~ ^2S_{1/2} - 5d ~ ^2D_{3/2}$ transition of $^{133}$Cs. Also, it may
be advantageous to consider either the $6s ~ ^2S_{1/2} (F=3) - 5d ~ ^2D_{5/2}(F=2;3;4)$ or $6s ~ ^2S_{1/2} (F=4)- 5d ~ ^2D_{5/2} (F=3,4,5)$ 
transitions to measure PNC effects or among all possible transitions to obtain an average value for reducing major systematic 
uncertainties. 

\begin{table}[t]
\caption{
Reduced E1 matrix elements in atomic unit (a.u.), $H_{PNC}^{NSI}$ matrix elements (in 
$-iea_{0} (Q_{W}/N)\times 10^{-11}$ and {\bf K} matrix elements in $-iea_{0} K_{W}\times 10^{-11}$ 
in the top part of the table. Wavelengths (in nm) and reduced E2 (in a.u.) and M1 (in a.u.) matrix elements 
in the bottom part of the table.
}
\begin{tabular}{lccc}
\hline
 \hline 
 $f - i$ transition &   E1 & $\langle J_f||{\bf H}_{PNC}^{NSI}||J_i\rangle$&$\langle J_f||{\bf K}||J_i  \rangle$  \\
 \hline 
& & & \\
 $6p ~ ^2P_{1/2} - 6s ~ ^2S_{1/2}$ & $-4.53$ & $-0.94$& $-2.30$   \\
 $7p ~ ^2P_{1/2} - 6s ~ ^2S_{1/2}$ & $-0.31$ &  $-0.55$ & 1.39   \\
 $8p ~ ^2P_{1/2} - 6s ~ ^2S_{1/2}$ & $-0.10$ &  $-0.37$ & 0.95   \\
 $9p ~ ^2P_{1/2} - 6s ~ ^2S_{1/2}$ & $-0.05$ &  $-0.27$ & 0.71   \\
 $10p ~ ^2P_{1/2} - 6s ~ ^2S_{1/2}$ & $-0.03$ & $-0.14$&  0.56   \\
 
 $6p ~ ^2P_{3/2} - 6s ~ ^2S_{1/2}$ & 6.40 &  & $0.26$   \\
 $7p ~ ^2P_{3/2} - 6s ~ ^2S_{1/2}$ & $-0.62$ &   & $-0.09$   \\
 $8p ~ ^2P_{3/2} - 6s ~ ^2S_{1/2}$ & $-0.25$ &   & $-0.05$   \\
 $9p ~ ^2P_{3/2} - 6s ~ ^2S_{1/2}$ & $0.14$ &   & 0.03   \\
 $10p ~ ^2P_{3/2} - 6s ~ ^2S_{1/2}$ & $0.09$ &  &  0.02   \\
  
 $5d ~ ^2D_{3/2} - 6p ~ ^2P_{1/2}$ & $-7.25$ &      &  $-0.24$   \\
 $5d ~ ^2D_{3/2} - 7p ~ ^2P_{1/2}$ & 2.29 &      &  $-0.04$  \\
 $5d ~ ^2D_{3/2} - 8p ~ ^2P_{1/2}$ & $0.70$ &      &  $-0.02$   \\
 $5d ~ ^2D_{3/2} - 9p ~ ^2P_{1/2}$& $0.38$ &      &  $-0.008$   \\
 $5d ~ ^2D_{3/2} - 10p ~ ^2P_{1/2}$& $0.26$ &      &  $-0.004$   \\
 
 $5d ~ ^2D_{3/2} - 6p ~ ^2P_{3/2}$ & $3.27$ &$-0.05$ &  $-0.05$   \\
 $5d ~ ^2D_{3/2} - 7p ~ ^2P_{3/2}$ & 0.91 & $0.03$ &  $0.08$   \\
 $5d ~ ^2D_{3/2} - 8p ~ ^2P_{3/2}$ & 0.28 & $0.02$ &  $0.05$   \\
 $5d ~ ^2D_{3/2} - 9p ~ ^2P_{3/2}$& $-0.16$ & $-0.02$ &  $-0.04$   \\
 $5d ~ ^2D_{3/2} - 10p ~ ^2P_{3/2}$& $-0.10$ & $-0.01$ &  $-0.03$   \\
 
 $5d ~ ^2D_{5/2} - 6p ~ ^2P_{3/2}$ & 9.92 &  &  $0.58$   \\
 $5d ~ ^2D_{5/2} - 7p ~ ^2P_{3/2}$ & 2.12 &   &  $-0.24$   \\
 $5d ~ ^2D_{5/2} - 8p ~ ^2P_{3/2}$ & 0.68 &   &  $-0.15$   \\
 $5d ~ ^2D_{5/2} - 9p ~ ^2P_{3/2}$& $-0.36$ &   &  $0.11$   \\
 $5d ~ ^2D_{5/2} - 10p ~ ^2P_{3/2}$& $-0.22$ &  &  $0.08$   \\
\hline 
 & & & \\
  $f - i$ transition & $\lambda_{fi}$ & E2 & M1  \\
 \hline 
  &  &  &  \\
 $5d ~ ^2D_{3/2} - 6s ~ ^2S_{1/2}$ & 689.69        &   34.38  & $\sim 0.0$  \\  
 $5d ~ ^2D_{5/2} - 6s ~ ^2S_{1/2}$  & 685.08        &   48.50  &       \\
 $5d ~ ^2D_{5/2} - 5d ~ ^2D_{3/2}$  & 102469.52        &   34.75  & 1.511   \\
\hline \hline
\end{tabular}
\label{tab3}
\end{table}

Alternatively, we can consider trapped $^{133}$Cs atoms in an optical lattice interacting with an oscillating electric field 
given by 
\begin{eqnarray}
\label{eq13}
\mbox{\boldmath${\cal E}$} ({\bf r},t) &=& \frac {1}{2} \left[ \mbox{\boldmath${\cal E}$} ({\bf r}) e^{-i\omega t} + c.c. \right],
\end{eqnarray}
where $\omega$ is the laser frequency and $\mbox{\boldmath${\cal E}$}$ is the strength of the electric field. Following the principle discussed in Ref. \cite{fortson}, the Rabi frequencies due to 
PNC and E2 amplitudes of the $6s ~ ^2S_{1/2} - 5d ~ ^2D_{3/2;5/2}$ transitions can be given by
\begin{eqnarray}
{\rm{\Omega}}^{M M'}_{PNC} &=& - \frac{1}{2\hbar} \sum_{i} \left( E1^{M M'}_{PNC} \right)_i {\cal E}_i(0),
\label{eq14}
\end{eqnarray}
and
\begin{eqnarray}
{\rm{\Omega}}^{M M'}_{E2} &=& - \frac{1}{2\hbar} \sum_{i,j} \left(E2^{M M'} \right)_{ij}
\left [ \frac{\partial{ {\cal E}_i({\bf r})}} {\partial{x_j} } \right ]_{r=0},
\label{eq15}
\end{eqnarray}
where $E1^{M M'}_{PNC}$ are either $E1_{PNC}^{NSI}$ or $E1_{PNC}^{NSD}$ or total PNC amplitude depending on whether 
$5d ~ ^2D_{3/2}$ or $5d ~ ^2D_{5/2}$ state is being considered in the transition and $E2^{M M'}$ is the corresponding E2 transition amplitude. The 
Rabi frequency (${\rm{\Omega}}^{M M'}$) due to interference between both the forbidden transition amplitudes would yield
\begin{eqnarray}
|{\rm{\Omega}}^{M M'}|^2 &=& |{\rm{\Omega}}^{M M'}_{E2} + {\rm{\Omega}}^{M M'}_{PNC}|^2 \nonumber \\
  & \simeq & |{\rm{\Omega}}^{M M'}_{E2}|^2 + 2Re \left( {\rm{\Omega}}^{M M'}_{PNC*} {\rm{\Omega}}^{M M'}_{E2} \right).
\end{eqnarray}
Obviously the contribution from the Rabi frequency due to E2 transition will be much larger than the detuning frequency. Using the 
above quantities, the light shifts due to PNC and E2 transitions are given by
\begin{eqnarray}
\label{eq16}
\Delta\omega^M_{PNC} &\approx& - \frac{Re \sum_{M'} \left({\rm{\Omega}}^{M M'}_{PNC*} {\rm{\Omega}}^{M M'}_{E2}\right)} 
{\sqrt{\sum_{M'}|{\rm{\Omega}}^{M M'}_{E2}|^2}}
\end{eqnarray}
and
\begin{eqnarray}
\Delta\omega^{M}_{E2} &\approx& \frac{(\omega_0 - \omega)}{2} - \sqrt{\sum_{M'}|{\rm{\Omega}}^{M M'}_{E2}|^2},
\label{eq17}
\end{eqnarray}
where $\omega_0$ is the resonant frequency of the considered transition before applying the laser field. 
This approximation is valid when the laser frequency $\omega$ is close to the resonance. By extracting 
${\rm{\Omega}}^{M M'}_{PNC}$ value for a set of azimuthal quantum numbers $M$ and $M'$ from the measurements 
of $\Delta \omega^M_{PNC}$ and $\Delta\omega^{M}_{E2}$ values, the $E1^{M M'}_{PNC}$ value can be extracted.
By combining $E1^{M M'}_{PNC}$ and the calculated values of $E1^{M M'}_{PNC}/Q_W$ or $E1^{M M'}_{PNC}/K_W$ value, it is 
possible to extract the values of $Q_W$ or $K_W$. In this paper, we give values of $E1^{M M'}_{PNC}$ by considering $M=M'$ as the 
smaller value among the hyperfine angular momenta of the initial state $F_i$ and the final state $F_f$ involved in the 
transition by substituting values of $Q_W$ and $K_W$ from the empirical relations. This would enable the experimentalists to choose suitable 
transitions among the hyperfine levels of the ground and $5d ~ ^2D_{3/2;5/2}$ states to measure the NSI and NSD PNC effects. For this purpose, we perform 
calculations of the $\mathcal{X}$ and $\mathcal{Y}$ values and discuss on them in the next section along with the other relevant 
calculations on spectroscopic properties that will be useful for analyzing uncertainties in the PNC results.

\begin{table}[t]
\caption{Demonstration of trends of the calculated EA (in cm$^{-1}$) of the first few low-lying excited states in Cs using DF and 
CCSD methods. Results using DC, DC along with individual corrections as Breit (DC$+$Breit), VP (DC$+$VP), SE (DC$+$SE) and all the 
corrections together (DC$+$all) are given systematically using the CCSD method. Estimated EE values from EAs are given below using the 
DF method and DC$+$all approximation in the CCSD method. These values are compared with the experimental results listed in the 
NIST database and with the calculated values (without scaling) reported in Ref. \cite{roberts}.}
\begin{ruledtabular}
\begin{tabular}{lccccc} 
 Method  & $6s ~ ^2S_{1/2}$ & $6p ~ ^2P_{1/2}$ & $6p ~ ^2P_{3/2}$ & $5d ~ ^2D_{3/2}$ & $5d ~ ^2D_{5/2}$ \\
 \hline
    & & & & & \\
   \multicolumn{6}{c}{\underline{EA values}} \\  
 DF & 27921.72 & 18789.33 & 18387.67 & 14118.17 & 14143.04 \\
 DC(CCSD) & 31356.64 & 20194.75 & 19635.76 & 16596.14 & 16504.14 \\
 DC$+$Breit & 31356.24 & 20187.25  & 19634.44 & 16616.31 & 16527.76 \\
 DC$+$VP & 31360.27 & 20194.72 & 19635.67 & 16595.74 & 16503.78 \\
 DC$+$SE & 31338.72 & 20193.66 & 19636.71 & 16598.25 &  16506.29 \\
 DC$+$all & 31341.78 & 20186.13 & 19635.29 & 16618.01 & 16529.56 \\
   & & & & & \\
Ref. \cite{roberts}$^{\ddagger}$ & 31457 & 20290 & 19722 & 17146 & 17030 \\
 NIST \cite{nist} & 31406.47 & 20228.20 & 19674.16  & 16907.21 & 16809.63 \\
 \hline
   & & & & & \\
    \multicolumn{6}{c}{\underline{EE values}} \\
 DF & 0.0 & 9132.39 & 9534.05 & 13803.55 & 13778.68 \\
 CCSD & 0.0 & 11155.65 & 11706.49 & 14723.77 & 14812.22 \\
   & & & & & \\
 NIST \cite{nist} & 0.0 & 11178.27 &  11732.21 & 14499.26 & 14596.85 \\
\end{tabular}
\end{ruledtabular}
\label{tab4}
$^{\ddagger}$Obtained using the DC Hamiltonian.
\end{table}

\begin{table*}[t]
\caption{Reduced matrix elements of E1, $H_{PNC}^{NSI}$ and {\bf K} operators from the DF and CCSD methods. CCSD calculations with 
different approximated Hamiltonians as mentioned in the previous table are given explicitly. Units of these results are same with the 
respective values quoted in Table \ref{tab3}. We also compare our E1 matrix elements with the calculated values (without scaling) 
reported in Ref. \cite{roberts}.}
\begin{ruledtabular}
\begin{tabular}{lccccc} 
  Method  &  $6p ~ ^2P_{1/2} - 6s ~ ^2S_{1/2}$ & $6p ~ ^2P_{3/2} - 6s ~ ^2S_{1/2}$ &  $5d ~ ^2D_{3/2} - 6p ~ ^2P_{1/2}$  & $5d ~ ^2D_{3/2} - 6p ~ ^2P_{3/2}$  & $5d ~ ^2D_{5/2} - 6p ~ ^2P_{3/2}$ \\
 \hline
    & & & & & \\
  \multicolumn{6}{c}{\underline{Reduced E1 matrix elements}} \\
  DF & 5.272 & 7.416 & 9.012 & 4.078 &  12.233 \\
 DC(CCSD) & 4.529 & 6.401 & 7.259 & 3.275 & 9.940 \\
 DC$+$Breit & 4.529 & 6.401 & 7.251 & 3.272 & 9.921 \\
 DC$+$VP &  4.529 &  6.400 &  7.260 & 3.275 & 9.940 \\
 DC$+$SE & 4.532 &  6.405 & 7.258 & 3.269 & 9.938 \\
 DC$+$all & 4.531 & 6.404 & 7.250 & 3.271 & 9.922 \\
   & & & & & \\
Ref. \cite{roberts}$^{\ddagger}$ & 4.506 & 6.343 & 6.926 & 3.121 & 9.481 \\
Experiment & 4.5097(74) \cite{young} & 6.3403(64) \cite{young} & 7.33(6) \cite{diberardino} & 3.28(3) \cite{diberardino} & 9.91(3) \cite{diberardino} \\
   & & & & & \\
   
 \multicolumn{6}{c}{\underline{Reduced matrix elements of $H_{PNC}^{NSI}$}} \\
 DF &  0.670 &  &  & $\sim 0.0$ & \\
 DC(CCSD) & 0.962 &  &  & 0.049 &  \\
 DC$+$Breit & 0.961 & &  & 0.048 & \\
 DC$+$VP & 0.966         & &  & 0.049 & \\
 DC$+$SE & 0.940   &  &  & 0.047 & \\
 DC$+$all & 0.941 & & & 0.047 & \\

    & & & & & \\ 
  \multicolumn{6}{c}{\underline{Reduced matrix elements of {\bf K}}} \\  
 DF & 2.003 & 0.0 &  0.0 & $\sim 0.0$ & 0.0\\
 DC(CCSD) & 2.352 & 2.267 & 0.242 & 0.048 & 0.589 \\
 DC$+$Breit & 2.346 & 0.266 & 0.243 & 0.049 & 0.592 \\
 DC$+$VP & 2.360 & 0.269 & 0.243 & 0.049 & 0.592 \\
 DC$+$SE & 2.300 & 0.260 & 0.235 & 0.048 & 0.575 \\
 DC$+$all & 2.300 & 0.260 & 0.237 & 0.049 & 0.579 \\
\end{tabular}
\end{ruledtabular}
\label{tab5}
$^{\ddagger}$Obtained using the DC Hamiltonian.
\end{table*}

\section{Results and Discussion}

In Table \ref{tab2}, we present $E1_{PNC}$ amplitudes for transitions between different allowed hyperfine levels, in accordance with the 
selection rules of Eq. (\ref{eq4}), with nuclear spin $I=7/2$ of the $6s ~ ^2S_{1/2} - 5d ~ ^2D_{3/2}$ and $6s ~ ^2S_{1/2} - 5d ~ ^2D_{5/2}$ transitions 
from the NSI and NSD interactions, given as $E1_{PNC}^{NSI}$ and $E1_{PNC}^{NSD}$ respectively. Here we use $q=0$ and the azimuthal 
quantum number of the hyperfine levels as the minimum value among the hyperfine angular momenta. We present these values 
from both the DF and CCSD methods to highlight the role of correlation effects. We have used the
values $Q_W=-73.2$ and $K_W=0.419$ for $^{133}$Cs that are determined using the relations \cite{ginges,flam1}
\begin{eqnarray}
Q_W &\simeq& -N + Z(1-\sin^2{\Theta_W} ) 
\end{eqnarray}
and
\begin{eqnarray}
 {\cal K}_W \approx \frac{9}{10} g_p \mu_p \frac{\alpha_e A^{2/3}}{M_p r_0},
\end{eqnarray}
where $\Theta_W$ is the Weinberg angle, $\alpha_e$ is the fine structure constant, $g_p \simeq 5.0$ is the nucleon-nucleon
parity-odd coupling and $\mu_p \simeq 2.8$ is the magnitude of the magnetic moment of the proton, $M_p$ is the proton mass and 
$r_0 \simeq 1.2$ fm.
 
The matrix elements that are used to determine the ``Main'' contributions to the $E1_{PNC}^{NSI}$ and $E1_{PNC}^{NSD}$ amplitudes are 
given in Table \ref{tab3}. We use the experimental energies given in the National Institute of Science and Technology (NIST) database 
\cite{nist}, along with these matrix elements for the sum-over-states approach. It is found from the differences between the ``Main'' 
contributions, which can be estimated using the above matrix elements, and the final results for the $E1_{PNC}^{NSI}$ and $E1_{PNC}^{NSD}$ amplitudes  
given in Table \ref{tab2} that both the ``Tail'' and ``relaxation'' contributions together are very small. We also find that the correlation 
contributions are very large for $E1_{PNC}^{NSD}$ corresponding to the different hyperfine levels of the $6s ~ ^2S_{1/2} - 5d ~ ^2D_{5/2}$ 
transition. We also compare our results with another calculation that are obtained employing a relativistic many-body perturbation 
theory \cite{roberts}. There is good agreement between the results for the $6s ~ ^2S_{1/2} - 5d ~ ^2D_{3/2}$ transition, while the 
NSD values from both the works differ significantly for the $6s ~ ^2S_{1/2} - 5d ~ ^2D_{5/2}$ transition. It is mentioned in Ref. \cite{roberts} that the results need to be improved further 
by including the Breit and QED interactions as well as higher-order non-Brueckner electron correlation effects. In fact, Ref. \cite{roberts} 
also uses a scaling procedure to estimate some missing correlation effects. Our RCC method takes into account both the core-polarization 
and different kinds of pair correlation effects to all orders. Corrections due to the Breit and lower order QED interactions are investigated explicitly 
as discussed below. It is worth mentioning here is that there are huge differences between the DF and CCSD results for the NSD 
interaction for the $6s ~ ^2S_{1/2} - 5d ~ ^2D_{5/2}$ transition and also the signs for both the results are opposite. This is because of the 
unusually large core-polarization effects associated in the calculations of the $5d ~ ^2D_{5/2}$ state. It appears that this is the main reason 
behind the large discrepancies between our results and the values reported in Ref. \cite{roberts}. Accuracies of our 
results can be improved further by taking recourse to an approach used by us in Refs. \cite{wansbeek,bijaya5} in the RCC framework.

It is obvious from Eqs. (\ref{eq5}) and (\ref{eq6}) that the precise determination of the values of $\mathcal{X}$ and $\mathcal{Y}$ to evaluate
$E1_{PNC}$ amplitudes require accurate calculations of the E1 matrix elements, EEs and matrix elements of the PNC interaction 
Hamiltonians. Therefore, we also discuss the trends of the contributions to these quantities from the correlation effects and relativistic 
corrections for a few important low-lying states that are significant for the calculations of $\mathcal{X}$ and $\mathcal{Y}$.
Accuracies of our calculated EEs in Cs can be gauged by comparing them with the available data in the NIST database \cite{nist}.
To find the accuracies of the calculated E1 matrix elements, we would like to estimate errors in the determination of lifetimes of the 
low-lying excited $6p ~ ^2P_{1/2;3/2}$ and $5d ~ ^2D_{3/2;5/2}$ states by substituting these values. It is, however, not possible to directly evaluate the accuracies of the PNC matrix elements  from any measured quantity, but they
could be indirectly estimated from the errors in the calculations of another quantity, i.e. hyperfine interaction constant, which like the PNC
interactions is also sensitive to the region near the nucleus. We have therefore calculated the magnetic dipole hyperfine structure 
constants ($A_{hyf}$) for the states of interest to us and compared them with their experimental values to assess their accuracies \cite{bks} . 

In Table \ref{tab4}, we present the EA of the ground state and EEs of a few low-lying excited states at the DF and relativistic CCSD 
levels of approximation and compare them with the NIST data. This table clearly shows that the
DF values differ substantially from the experimental and relativistic CCSD values with different corrections that are very close to the
experimental results. Corrections from the Breit and QED interactions were found to be small but necessary for high precision results.
These corrections to the relativistic CCSD values in the DC approximation are determined by incorporating individual QED interactions 
step by step and then taking all the interactions together. Contributions from these approximations are given in the above mentioned 
table. It shows that both the EA and EEs agree with the measured values to less than one percent except for the the $5D$ states. 
Inclusion of triple excitations might improve the results for the $5D$ states. We also compare our results in the above table with the
values reported in Ref. \cite{roberts} using the DC Hamiltonian and find good agreement between them. 

\begin{table}[t]
\caption{Magnetic hyperfine structure constants (in MHz) of the first few low-lying states of $^{133}$Cs for the DF and different
CCSD approximations. We have used the nuclear gyromagnetic constant as $g_I=0.7379751$ to determine 
these values. We have also compared our calculations with the experimental values given in Refs. \cite{arimondo,rafac1,tanner}
and with the calculated values reported in Ref. \cite{roberts}.}
 \begin{ruledtabular}
  \begin{tabular}{lccccc}
 Method  &  $6s ~ ^2S_{1/2}$ & $6p ~ ^2P_{1/2}$ & $6p ~ ^2P_{3/2}$ & $5d ~ ^2D_{3/2}$ & $5d ~ ^2D_{5/2}$  \\
  \hline
 & & & & & \\
 DF & 1431.62 & 161.27 & 23.93 & 18.15 &  7.44 \\
 DC(CCSD) & 2304.39 & 284.91 & 48.65 & 46.31 & $-17.51$ \\
 DC$+$Breit & 2308.07 & 284.90 & 48.75 & 46.80 & $-17.54$ \\
 DC$+$VP & 2309.01 & 284.97 & 48.66 & 46.35 & $-17.55$ \\
 DC$+$SE & 2280.23 & 284.24 & 48.60 & 46.14 & $-17.30$ \\
 DC$+$all & 2288.08 & 284.28 & 48.72 & 46.66 & $-17.37$ \\
   & & & & & \\
 Ref. \cite{roberts}$^{\ddagger}$ & 2315 & 290 &  &  &  \\
 Experiment & 2298 & 291.89(8) & 50.275(3) & 48.78(7) & $-21.24(8)$\\
  \end{tabular}
 \end{ruledtabular}
 \label{tab6}
 $^{\ddagger}$Obtained using the DC Hamiltonian.
\end{table}
We also give the results for the matrix elements of the E1, $H_{PNC}^{NSI}$ and $K$ operators using the DF and relativistic CCSD methods 
along with the Breit and QED interactions at different levels of approximation for some of the dominant contributions to $E1_{PNC}$ 
in Table \ref{tab5}. These results show the dominant role of correlation effects for these quantities. Corrections from the Breit
and QED interactions are found to be relatively small in the evaluation of E1 matrix elements, but they make sizable contributions to the
PNC matrix elements; especially that of the self energy. We have also compared our results with the values obtained in Ref. \cite{roberts} 
using the DC Hamiltonian and without scaling the calculated wave functions. Considering the final CCSD values of E1, E2 and M1 matrix 
elements from the DC Hamiltonian along with Breit and QED interactions given in Table \ref{tab3}, we find that the lifetimes of the 
$6p ~ ^2P_{1/2}$,  $6p ~ ^2P_{3/2}$, $5d ~ ^2D_{3/2}$  and $5d ~ ^2D_{5/2}$ states are 34.44 ns, 29.85 ns, 917.5 ns and 1280.1 ns respectively. Our estimated 
lifetime value 34.44 ns for the $6p ~ ^2P_{1/2}$ state is in close agreement with two experimental values reported as 34.75(7) ns \cite{young} 
and 35.07(10) ns \cite{rafac}. Similarly our value 29.85 ns for the $6p ~ ^2P_{3/2}$ state also match reasonably with experimental values as
30.41(10) ns \cite{young} and 30.57(7) ns \cite{rafac}. We also obtain reasonable agreement between the measured lifetimes of the 
$5d ~ ^2D_{3/2}$ and $5d ~ ^2D_{5/2}$ states as 909(15) ns \cite{diberardino} and 1281(9) ns \cite{diberardino} respectively. For all the states, the
dominant contributions to the lifetimes are from the E1 matrix elements. The results of the calculated and measured values of the 
lifetimes are in good agreement, particularly for the $5d ~ ^2D_{3/2;5/2}$ states. This suggests that our E1 matrix elements are accurate enough for 
highlighting the importance of the $6s ~ ^2S_{1/2} - 5d ~ ^2D_{3/2;5/2}$ transition amplitudes for PNC studies. They can be further improved by
including the triple excitations in our calculations. In order to make error estimates for the $E1_{PNC}$ amplitudes, we also 
list the extracted E1 matrix elements from the above measured lifetimes as experimental values in Table \ref{tab3}.

We now present $A_{hyf}$ values in Table \ref{tab6} of the five low-lying states that are relevant in gauging the accuracies of the 
$H_{PNC}^{NSI}$ and $K$ matrix elements of the $6s ~ ^2S_{1/2} - 5d ~ ^2D_{3/2;5/2}$ transitions. Like the previous results, we also give these 
quantities at the DF and CCSD levels along with the Breit and QED interactions. We have used the nuclear gyromagnetic constant 
$g_I=0.7379751$ for the evaluation of the $A_{hyf}$ values. In the above table, we have also given the precisely known experimental 
results and they are in good agreement with our calculated values. The differences between the DF and CCSD results show that correlation 
effects contribute significantly and the DF and CCSD values have opposite signs for the $5d ~ ^2D_{5/2}$ state. We also find that both the 
Breit and QED corrections contribute negligibly to these results except for the ground state. We also compare our results with the 
values reported in Ref. \cite{roberts} using the DC Hamiltonian. We expect that accuracies in our results will be improved 
further by adding contributions from the triples excitations.

The magnitudes of the $E1_{PNC}$ values, given in Table \ref{tab2}, can help in 
identifying suitable hyperfine transitions in $^{133}$Cs for carrying out PNC measurements reliably. It can be seen that these values are 
large for the transitions with the same hyperfine angular momentum for the initial and final states. In fact, the $E1_{PNC}^{NSD}$ 
amplitudes in the $6s ~ ^2S_{1/2} - 5d ~ ^2D_{5/2}$ transition are found to be of similar order of magnitude as the 
$6s ~ ^2S_{1/2} - 5d ~ ^2D_{3/2}$  
transition for these cases. It would be judicious to first perform a PNC measurement for the $6s ~ ^2S_{1/2} - 5d ~ ^2D_{5/2}$ transition to 
extract precise information on the NSD interaction which is dominated by the NAM. Using this information, it would be possible to separate the contributions to the 
measured PNC effects for the $6s ~ ^2S_{1/2} - 5d ~ ^2D_{3/2}$ transition due to the NSD and NSI interactions. Different spectroscopic 
properties of Cs calculated by us suggest that our relativistic CCSD method is capable of giving accurate results. We quantify the
uncertainties associated with the $E1_{PNC}$ amplitudes using the errors from the individual quantities involved in determining them.
It can be seen from Eqs. (\ref{eq5}) and (\ref{eq6}) that uncertainties in $E1_{PNC}$ come from the errors associated with the E1 matrix 
elements, matrix elements of the PNC interaction Hamiltonians and the excitations energies between different states of the Cs atom. We
evaluate the errors in these quantities by comparing our calculated values with their corresponding experimental results. Since we have 
used experimental energies in our calculations, we only take into account errors from the E1 and PNC Hamiltonian matrix elements. We 
consider weighted contributions from the matrix elements between the $6s ~ ^2S_{1/2}$, $6p ~ ^2P_{1/2;3/2}$ and $5d ~ ^2D_{3/2;5/2}$ 
states, owing to their dominant 
contributions to determine the uncertainties in the $E1_{PNC}$ amplitudes. We estimate the errors in the matrix elements of the PNC Hamiltonians between two 
given states, say, $|\Psi_i\rangle$ and $|\Psi_f\rangle$, from the errors involved in the hyperfine structure constants 
by assuming that the error in $\langle \Psi_f |H_{PNC}| \Psi_i \rangle$ can be approximated by the error in $\sqrt{A_{hyf}^i A_{hyf}^f}$ with 
the hyperfine structure constants $A_{hyf}^i$ and $A_{hyf}^f$ of the states $|\Psi_i \rangle$ and $|\Psi_f\rangle$ respectively
\cite{bijaya5,bks}. By adding the errors in the E1 and $H_{PNC}$ matrix elements in quadrature,  we estimate the uncertainties 
in $E1_{PNC}^{NSI}$ and $E1_{PNC}^{NSD}$ amplitudes as approximately 2\% and 3\%  respectively for the $6s ~ ^2S_{1/2} - 5d ~ ^2D_{3/2}$ transition. Similarly, we find about 2\% 
uncertainty in the $E1_{PNC}^{NSD}$ amplitude of the $6s ~ ^2S_{1/2} - 5d ~ ^2D_{5/2}$ transition. It is possible to improve the accuracies 
of the theoretical uncertainties further by including higher level excitations in our RCC calculation. An experiment 
to measure PNC  in the $6s ~ ^2S_{1/2} - 5d ~ ^2D_{3/2;5/2}$ transitions in Cs would mark an important advance in the field of atomic PNC.

\section{Conclusion}

We have carried out a relativistic coupled-cluster analysis of the PNC effects for all possible hyperfine levels in 
the $6s ~ ^2S_{1/2} - 5d ~ ^2D_{3/2}$ and $6s ~ ^2S_{1/2} - 5d ~ ^2D_{5/2}$ transitions of $^{133}$Cs. We find that the PNC effects due to the NSD interaction in the 
$6s ~ ^2S_{1/2} - 5d ~ ^2D_{5/2}$ transition are of  similar magnitudes as those for the $6s ~ ^2S_{1/2} - 5d ~ ^2D_{3/2}$  transition and
proposals for measuring them by two different techniques have been briefly outlined. The results of our present work  provides 
useful insights into designing an experiment for the unambiguous observation of the nuclear anapole moment. 

This paper is to dedicated Professor Debashis Mukherjee on the occasion of his 70$^{th}$ birthday. We have had a very fruitful 
collaboration with him for the past two decades on the application of relativistic coupled cluster theory to parity and time-reversal 
violations in atoms.

\section*{Acknowledgement}

Computations reported in this work were performed using the PRL Vikram-100 HPC cluster.

\end{document}